# Machine learning the screening factor in the soft bond valence approach for rapid crystal structure estimation

Keisuke Kameda[a*], Takaaki Ariga[a], Kazuma Ito[a], Manabu Ihara[a*], and Sergei Manzhos[a*]

[a]Department of Chemical Science and Engineering, School of Materials and Chemical Technology, Tokyo Institute of Technology, 2-12-1 Ōokayama, Meguro-ku, Tokyo, 152-8552, Japan

Correspondence email (Sergei Manzhos): manzhos.s.aa@m.titech.ac.jp

Correspondence email (Manabu Ihara): mihara@chemeng.titech.ac.jp

Email (Keisuke Kameda): kameda.k.ac@m.titech.ac.jp

## Abstract

The development of novel functional ceramics is critically important for several applications, including the design of better electrochemical batteries and fuel cells, in particular solid oxide fuel cells. Computational prescreening and selection of such materials can help discover novel materials but is also challenging due to the high cost of electronic structure calculations which would be needed to compute the structures and properties of interest such as the material's stability and ion diffusion properties. The soft bond valence (SoftBV) approach is attractive for rapid prescreening among multiple compositions and structures, but the simplicity of the approximation can make the results inaccurate. In this study, we explore the possibility of enhancing the accuracy of the SoftBV approach when estimating crystal structures by adapting the parameters of the approximation to the chemical composition. Specifically, on the examples of perovskite- and spinel-type oxides that have been proposed as promising solid-state ionic conductors, the screening factor – an independent parameter of the SoftBV approximation – is modeled using linear and non-linear methods as a function of descriptors of the chemical composition. We find that making the screening factor a function of composition can noticeably improve the ability of the SoftBV approximation to correctly model structures, in particular new, putative crystal structures whose structural parameters are yet unknown. We also analyze the relative importance of nonlinearity and coupling in improving the model and find that while the quality of the model is improved by including nonlinearity, coupling is relatively unimportant. While using a neural network showed no improvement over linear regression, the recently proposed GPR-NN method that is a hybrid between a single hidden layer neural network and kernel regression showed substantial improvement, enabling the prediction of structural parameters of new ceramics with accuracy on the order of 1%.

## 1 Introduction

In the development of novel materials for various applications, computation-guided design has been acquiring increasing importance. The availability of methods to compute properties and the availability of significant and growing CPU resources in principle permit in-silico discovery of new promising materials before more expensive experimental work is engaged.[1–5] Computation-guided design is particularly important for functional ceramics needed in technologies such as electrochemical batteries, fuel cells, electrolysis cells, and other technologies important for sustainable energy generation, storage, and use.[6–10] This includes functional oxides for solid-state ionic applications: solid-state metal-ion batteries (SSB)[11,12] and solid oxide fuel cells/electrolysis cells (SOFC/SOEC),[13] where the development of novel solid-state ionic conductors for various ions (alkali and alkali earth metal ions for SSB, protons and oxide ions for SOFC/SOEC) is still needed that would possess sufficient ionic conductivity as well as thermodynamic and redox stability and sufficiently low cost.[14–17] All these applications have much in common: for all types of conducted ions, there is the similarity of conceptual frameworks that can be employed for their understanding and design, the similarity of promising types of materials for them, and the similarity of modeling methods that can be used to produce mechanistic insight and to computationally pre-screen and guide the experimental development of new materials. There are also differences due to different mechanisms of ion-host interactions with different conducted ions. There is a vast design space, in particular, for mixed and doped oxides, which likely contain efficient solid electrolytes. The challenge is getting to the right material in that space. Computational prescreening and mechanistic insight-directed search are ways to achieve this.

Density functional theory (DFT)[18,19] is in principle sufficiently accurate to ascertain the required properties of a ceramic material with a putative composition and (crystal) structure. It can provide mechanistic insights, control, and resolution not easily achievable experimentally, but the relatively high computational cost of DFT calculations makes prescreening of all conceivable structures, let alone all ionic conduction paths, in a wide range of candidate materials too tedious. Such prescreening can in principle be done at the force field level if a force field framework is available that can be used for a wide range of ceramics and provide sufficient accuracy without requiring refitting of the force field for every new composition and structure. Most promising material candidates can then be subject to more detailed analysis with DFT and ultimately experimental verification.

The soft bond valence approximation (SoftBV) developed by Adams and co-workers provides such a framework.[20–22] It is a type of two-body force-field approximation that incorporates assumptions about the physics of bonding interactions. It is based on the bond valence approximation[23] and the inclusion of screened Coulombic interactions, which is appropriate for sufficiently ionic bonding. In this approach, one introduces a Bond Valence Site Energy (BVSE) which is a sum of contributions from all cations *i* [21,22]

$$E_{BVSE} = \sum_{i=1}^{M} E_{BVSE,i} \tag{1}$$

$$= \sum_{i=1}^{M}\left[\sum_{j=1}^{N_j} D_{0,ij}\left(\left(\frac{s_{ij}}{s_{min,\,ij}}\right)^2 - \frac{2s_{ij}}{s_{min.\,ij}}\right) + \sum_{i'=1}^{i'\neq i}\frac{q_i q_{i'}}{R_{ii'}}\cdot erfc\left(\frac{R_{ii'}}{sf\cdot(r_i + r_{i'})}\right)\right]$$

where the sum over *j* is the sum over anions, $s_{ij}(R_{ij}) = \exp\left(\frac{R_{0,\,ij}-R_{ij}}{b_{ij}}\right)$ is bond valence at the interatomic distance of $R_{i,j}$ between *i* and *j* ions, $s_{min,\,ij} = s_{ij}|_{R_{ij}=R_{min,ij}}$ is the value of $s_{ij}$ at the "equilibrium" geometry described by interatomic distances $R_{min,ij}$. $D_{0,ij}$, $R_{0,\,ij}$, $b_{ij}$, and screening factor (*sf*) are parameters. $r_i$ are ionic radii. $q_i$ are effective charges of the ions. Here and in the following, we use indices *i* for cations and *j* for anions unless stated otherwise. The sum in Eq. (1) is taken over ion's $N_j$ nearest neighbors (typically first coordination sphere defined by a cutoff radius $R_{cutoff,ij}$ which is another parameter) and all cations whose number is *M*. The choice of summation as a function of the atomic environment gives it a flavor of a reactive force field. Coulombic interactions, contrary to common force fields, are only explicitly included for repulsion between effective charges $q_i$ (see below) and are screened (controlled by *sf*). In Eq. (1), strictly speaking, only *sf* is an unconstrained free parameter. Relations have been established among the other parameters. The parameters $b_{ij}$ can be expressed via ionic softness (inverse of hardness[24]) $\sigma$ of the anion *A* and cation *C*, $b_{ij} = \sum_{n=0}^{5} a_i\left(\sigma_j^{(A)} - \sigma_i^{(C)}\right)^n$ where $a_i$ are coefficients fitted based on empirical HSAB (hard and soft acids and bases) concept.[22] A consistent set of relations between parameters has been developed[21,22,25,26] by making the SoftBV force field agree with known structures and other known force fields such as a universal force field (UFF).[27] According to those works, the bond breaking energy $D_{0,ij}$ is related to $b_{ij}$ and the oxidation state $V_{i,j}$ as:[21,25]

$$D_{0,ij} = \kappa\frac{b_{ij}^2}{2}\frac{c\left(V_iV_j\right)^{1/c}}{R_{min,ij}\left(n_in_j\right)^{1/2}} \qquad (2)$$

where $\kappa$ is a coefficient ($\kappa = 14.4$ eV Å$^{-1}$ if these units are used), *c* is related to the maximum angular momentum of the valence shell of the cation (*c* = 1 for *s*- and *p*- block elements, and 2 for *d*- and *f*- block elements), and $n_i$, $n_j$ are the principal quantum numbers of the cation and anion.[21,25] $R_{0,\,ij}$ can be thought of as the bond length resulting between the anion and cation when the cation contributes one valence to the anion;[28] it is related to other parameters as[21,25,26]

$$R_{min,ij} = \left(\gamma_1 + \gamma_2\left|\sigma_i - \sigma_j\right|\right)R_{0,ij} - b_{ij}ln\left(\frac{V_i}{N_c}\right) \tag{3}$$

where $\gamma_{1,2}$ are coefficients and $N_C$ is the coordination number. When matching to UFF, there is also a relationship between $D_{0,ij}$, $R_{0,ij}$, and $b_{ij}$:[26]

$$D_{0,ij} \approx \kappa\frac{b_{ij}^2}{2}\frac{c\left(V_iV_j\right)^{\frac{1}{c}}}{R_{min,ij}\left(n_in_j\right)^{\frac{1}{2}}}R_{0,ij} \tag{4}$$

The effective charges $q_i$ and $q_j$ of anions and cations in Eq. (1) are typically calculated as[21,25]

$$q_i = \frac{V_i}{\sqrt{n_i}}\left(\frac{\sum_j\frac{V_jN_j}{\sqrt{n_j}}}{\sum_i\frac{V_iN_i}{\sqrt{n_i}}}\right)^{\frac{1}{2}}, q_j = \frac{V_j}{\sqrt{n_j}}\left(\frac{\sum_i\frac{V_iN_i}{\sqrt{n_i}}}{\sum_i\frac{V_jN_j}{\sqrt{n_j}}}\right)^{\frac{1}{2}} \tag{5}$$

This ensures, in particular, the overall charge neutrality. A relationship between $R_{cutoff,ij}$ and other parameters have also been proposed:[22]

$$R_{cutoff,ij} = R_{0,ij} - b_{ij}ln\left(\frac{s_{ij}\left(R_{cutoff\,ij}\right)}{k}\right) \tag{6}$$

where $k$ is an empirical coefficient. Ionic radii are typically preset to agree with the literature;[29,30] their sum in Eq. (1) is fully correlated with *sf*.

The SoftBV approach provides a measure of material's stability via the Global Instability Index (GII)[21]

$$GII = \left(\frac{1}{N}\sum_{i=1}^{N}\left(\sum\nolimits_j s_{ij} - V_i\right)^2\right)^{1/2} \tag{7}$$

where $V_i$ are the formal oxidation states and $N$ is the number of cations. It also provides the ability to quickly prescreen ion conduction properties as Eq. (1) provides a potential energy map. In particular, the availability of a Bond Valence Path analyzer (BVPA), that analyses the topology of $E_{BVSE}$ as a function of transiting ion position,[20] makes it easy to rapidly compute all conduction paths for a given ion in a material, which is instrumental for understanding the nature of the diffusion (1D, 2D, 3D) and rate-limiting diffusion events. The method has been shown to be efficient for the prescreening of conductors for cations such as $Li^+$,[31–34], $Na^+$,[35,36] $Mg^{2+}$,[37] and $Zn^{2+}$ [38] for SSB. The approximations made to achieve high-throughput screening inevitably limit the quantitative accuracy compared to DFT. For example, for metal cations conducted ions, while trends in diffusion barriers agree well with DFT, their values can differ on the order of 1 eV.[20] Protons and oxide ions (of interest to SOFC/SOEC) are more challenging, in particular, as their interactions with the host are less ionic, and the two-body approximation and the simple expression of Eq. (1) are less reliable.[39]

SoftBV is often used for fixed crystal structures. Comparisons of properties (site energies, diffusion paths, etc.) at any level of theory are only meaningful if the structure is known with sufficient accuracy. For materials with new, putative compositions, optimal structures are unknown. It is desirable to have sufficient force field accuracy to find the correct structure directly with SoftBV without engaging in much more expensive DFT calculations or experiments. The ability to predict the structure would facilitate using more accurate methods (such as DFT) for energetic analysis, as the cost of optimization is then saved. It is in principle possible to improve the accuracy of the SoftBV approximation by adjusting its parameters, for example by making them depend on the composition or chemical environment of the atom. While $b_{ij}$, $D_{0,ij}$, $R_{0,ij}$, $R_{cutoff}$ or $N_C$, and charges still can be treated as tunable parameters and made depending on the chemical environment (see e.g. Ref. [40]), it would be at a cost of tempering with the basis of SoftBV ideology unless restrictions are imposed enforcing interrelations between the parameters such as those indicated above. This issue does not arise when tuning or parameterizing *sf*. When the structure of a material *is* known, *sf* can be automatically set to minimize the pressure, thus effectively tuning *sf* to the structure (lattice constants).[21] This value will in the following be called $sf_{auto}$. When prescreening for new materials with putative compositions where the optimal (correct) structure is not known, this approach in principle results in a non-optimal value of *sf* (i.e. in a $sf_{auto}$ value optimal for a wrong structure).

In this study, we therefore aim to determine an optimal value of the screening factor when the structure is not known, as a function of composition. We use linear and neural network (NN) models and show, on the examples of perovskite- [41–43] and spinel-type oxides[44] which have been proposed as promising solid-state ionic conductors, that this can noticeably improve the ability of the SoftBV approximation to model structures, in particular new, putative crystal structures whose structural parameters are yet unknown. We show that due to the smallness of the training dataset, there is no improvement with a neural network over the linear regression in spite of the higher expressive power of an NN. We employ a recently proposed machine learning method (called in the following GPR-NN) that is a hybrid between a neural network and kernel regression; in particular, it avoids nonlinear parameter optimization that is a cause of overfitting. GPR-NN allows building optimal nonlinear functions and controlling the inclusion of coupling between the features,[45] to analyze the importance of nonlinearity and of coupling and find that while the quality of the model is improved by including nonlinearity, the coupling is relatively unimportant. Overall, GPR-NN allowed the most accurate estimation of the optimal screening factor as a function of composition.

## 2 Methods

We fit $sf$ as a function of other SoftBV parameters that carry the information about the chemical composition ($b_{ij}$, $R_{0,ij}$, $R_{cutoff,ij}$, $r_i$, and $N_C$, which thus form the feature space). These features are available during SoftBV calculations. We consider 115 perovskite-type oxides with a general formula $ABO_3$ and 128 spinel-type oxides with a general formula $AB_2O_4$ where A and B are cations. These crystal structures are shown in Figure 1. The list of all materials is given in the Supplementary Material. These structures are taken mostly from Materials Project[46] and several from the ICDD database.[47] The structures taken from ICDD were confirmed by DFT calculations in Quantum Espresso[48] (using PBE[49] functional PAW pseudopotentials, and a plane wave cutoff of 35 Ry).

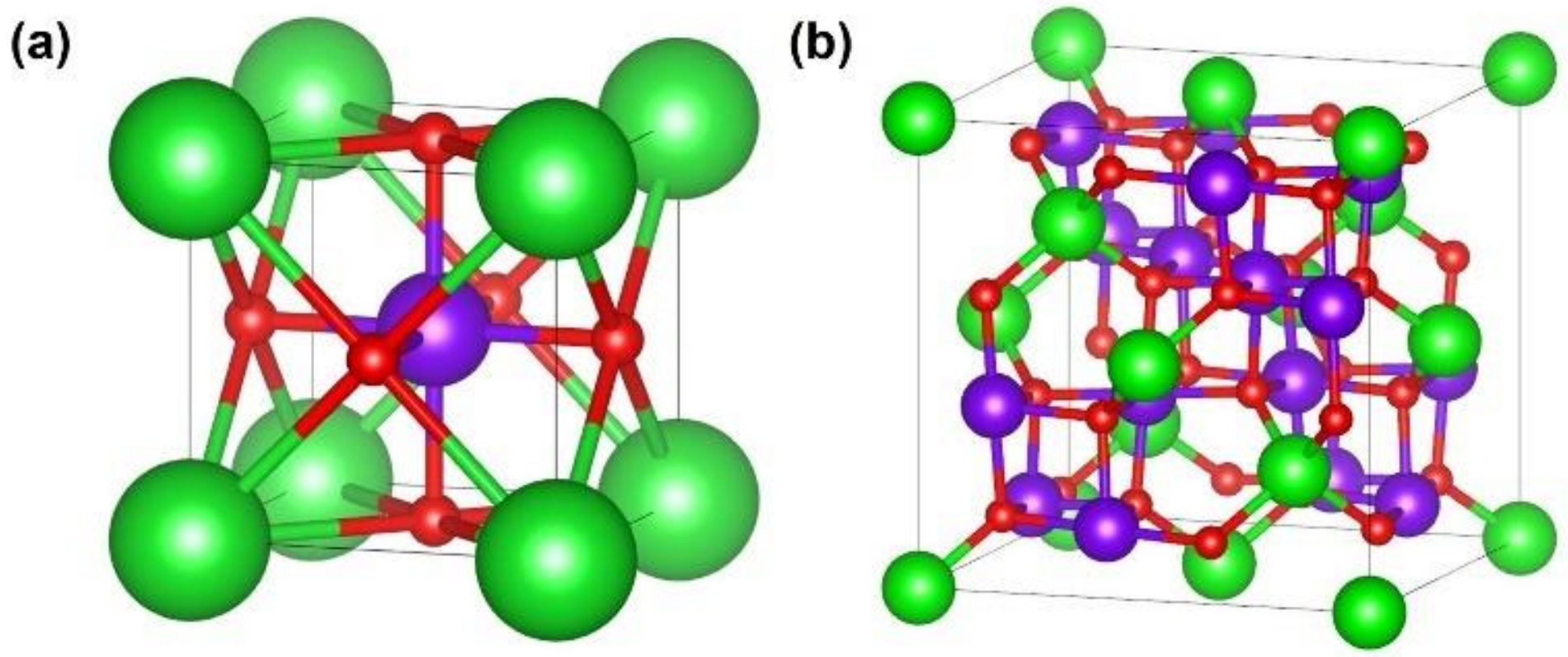

Figure 1. Crystal structures of (a) perovskite (A – green, B – violet, O - red), (b) spinel (A – green, B – violet, O – red) oxides.

Considering the relatively large dimensionality of the feature space, the number of data points (number of structures) is small.[50] We, therefore, perform the following procedure as shown in Figure 2: from each real structure obtained from the database called reference structure in the following, we form structures with lattice vectors isotopically expanded or contracted by 10%; these are called sample structures in the following. $sf$ is then set to values from 0.55 to 0.75 at 0.0125 intervals and structure optimization was performed in SoftBV. The error $Er$ – the difference between the lattice constants (defined below) following SoftBV optimization – is then collected resulting in a dataset of $b_{ij}$, $R_{0,ij}$, $R_{cutoff,ij}$, $r_i$, $N_C$, $sf$, and $Er$ for each reference or sample structure. In this way, the number of data points is expanded severalfold. In the case of perovskite-type oxides, SoftBV optimization does not result in any changes in fractional positions of atoms or distortions of the rectilinearity of the unit cell, and $Er$ is defined as the mean relative error in lattice vectors $a = b = c$ (i.e. $Er = (a_{reference} - a_{sample})/a_{reference} = (b_{reference} - b_{sample})/b_{reference} = (c_{reference} - c_{sample})/c_{reference}$). In the case of spinel-type oxides, SoftBV optimization results in small changes in the fractional positions of atoms within the unit cell. We defined the changes in fractional position per number of ions ($N$) as $\Delta_{\text{site}} = 1/N\sum_{i=1}^{N}\sqrt{(\Delta x^2 + \Delta y^2 + \Delta z^2)^2}$, where $N$ is the number of atoms in the cell and $\Delta x$, $\Delta y$, and $\Delta z$ are errors in fractional coordinates. $\Delta_{\text{site}}$ was lower than 0.01 in most of the spinel-type oxides, i.e. the error in structural parameters is mostly due to the lattice constants. Therefore, *Er* defined above was also used for optimizing crystal structures of the spinel-type oxides.

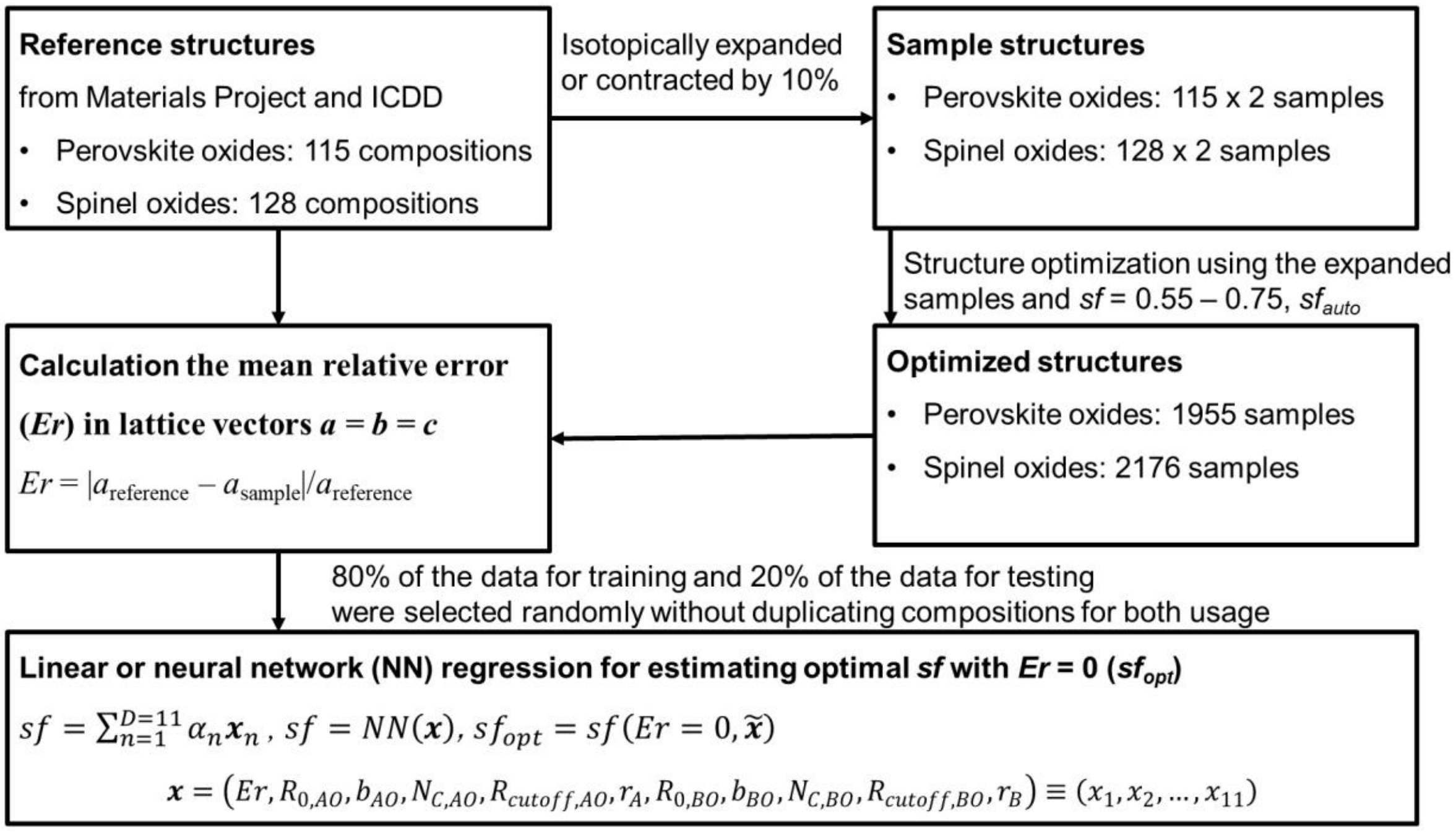

Figure 2. The procedure of optimizing the screening factor.

We define a $D$ = 11 dimensional vector of descriptors

$$\boldsymbol{x} = \left(Er, R_{0,AO}, b_{AO}, N_{C,AO}, R_{cutoff,AO}, r_A, R_{0,BO}, b_{BO}, N_{C,BO}, R_{cutoff,BO}, r_B\right)$$
$$\equiv (x_1, x_2, \ldots, x_{11})$$

and $\tilde{\boldsymbol{x}}$ as a vector of all descriptors other than $Er$. The dataset of ($\boldsymbol{x}$, *sf*) values for all materials and all structure expansions/contractions used for machine learning is provided in Supplementary Material. Ionic radii (which are coordination-dependent in SoftBV) for the coordination number of 6 were used in all cases. 80% of the expanded sample data of $\boldsymbol{x}$ and $sf$ randomly selected without duplicating compositions for training and testing were used for the training of the following regression models, and the remaining 20% for the testing. The features ($\boldsymbol{x}$) are normalized before fitting (i.e. its average and standard deviation are set to 0 and 1, respectively).

We perform linear regressions using the "*regress*" function in MATLAB:

$$sf = \sum_{n=1}^{D=11} \alpha_n x_n \tag{8}$$

We also perform non-linear regression using a feed-forward neural network (NN):[51]

$$sf = NN(\boldsymbol{x}) \tag{9}$$

The NN regressions are performed in MATLAB using "*trainlm*" function. Levenberg-Marquardt algorithm[52] was used to train the NN. We considered different numbers of hidden layers and neurons. *"tansig"* neuron activation function is used in the following. Other neuron activation functions were tried but resulted in no improvement (not shown). The estimated optimal *sf* ($sf_{est}$) was obtained from Eqs. (8) – (10) by setting *Er* = 0, i.e. $sf_{opt} = NN(0,\ \widetilde{\boldsymbol{x}})$. SoftBV optimization of crystal structures with expanded or contracted lattice was carried out using $sf_{auto}$ and $sf_{opt}$, and the *Er* was compared to evaluate the accuracy of SoftBV.

For the analysis of the relative importance of nonlinearity and coupling among the features, we use the GPR-NN method of Manzhos and Ihara.[45] The reader is referred to Refs. [45,53,54] for more details and context; here, we only briefly summarize the key properties of the method relevant to the purpose of the present work. The target function $sf(\boldsymbol{x})$ is expressed as

$$sf(\boldsymbol{x}) = \sum_{n=1}^{N} f_n(\boldsymbol{w}_n\boldsymbol{x}) = \sum_{n=1}^{N} f_n(y_n(\boldsymbol{x})) \tag{10}$$

This is a first-order additive model in (generally) redundant coordinates $\boldsymbol{y} = \boldsymbol{W}\boldsymbol{x}$, where $\boldsymbol{W}$ is the matrix of coefficients. The rows of $\boldsymbol{W}$ are defined as elements of a *D*-dimensional

Sobol sequence[55] although other ways of setting $\boldsymbol{W}$ are possible.[45] The shapes of the functions $f_n$ are computed using the first-order additive GPR[53,54,56–58] in $\boldsymbol{y}$. They are optimal for given data and given $\boldsymbol{W}$ in the least squares sense.[56] The original coordinates $\{x_n\}$ are also included in the set of $\{y_n\}$. If only $\{x_n\}$ are included, the method defaults to first-order additive GPR.[45,56,57] The representation of Eq. (10) is equivalent to a single hidden layer NN with optimal and individual to each neuron activation functions, and with weights fixed by rules rather than optimized. Matrix $\boldsymbol{W}$ is equivalent to the matrix of NN weights, while biases are subsumed in the definition of $f_n$. One can say that Eq. (10) is an NN in $\boldsymbol{x}$ and a 1st-order additive GPR in $\boldsymbol{y}$. The method has the advantage that because no nonlinear optimization is done, it does not suffer from overfitting as the number of 'neurons' $N$ grows beyond optimal,[45] combining the high expressive power of an NN and the robustness of linear regression (with nonlinear basis functions) which is GPR.[59] In this work, we use an additive RBF kernel in $\boldsymbol{y}$: $K(\boldsymbol{y}, \boldsymbol{y}') = \sum_{n=1}^{N} k(y_n, y_n')$ where $k(y_n, y_n') = exp\left(-\frac{(y_n - y_n')^2}{2l^2}\right)$. The data are normalized so that an isotropic kernel is used with a single length parameter $l$. In this work, we use this method to probe the importance of coupling terms by testing different $N$. In the limit of large $N$ the model fully includes all coupling among features, while in the limit $\boldsymbol{y} = \boldsymbol{x} \in R^D$, no coupling is included. On the other hand, the construction of optimal shapes of $f_n$ in the method is used to study the importance of nonlinearity. Similar to the case of an NN fit, $sf_{opt}$ is computed from the model of Eq. 10 by setting $Er = 0$, i.e. $sf_{opt} = f(0, \tilde{\boldsymbol{x}})$.

# 3 Results and discussion

## *3.1 Machine learning the screening factor with linear regression and neural networks*

Figure 3 shows the relationship between *Er* and *sf*. *Er*, namely the error in the lattice parameter, increased with an increase in *sf*. A larger *sf* makes the Coulombic repulsion in BVSE stronger at long range as per Eq. (1). Because the stress in a given crystal structure is to a significant degree due to Coulombic repulsion, SoftBV optimization with large *sf* resulted in an overestimated lattice constant. The relationship between *Er* and *sf* was different for each composition but did not depend on the initial lattice parameter. For instance, for two perovskite-type oxides of $BaCeO_3$ (orange squares) and $LaGaO_3$ (green triangle), one obtains *Er* = 0 with *sf* of about 0.60 and 0.65, respectively (Figure 3 (a) and (b)). Similar results were also obtained in the case of spinel-type oxides (i.e. $MnCo_2O_4$ (red squares) and $ZnFe_2O_4$ (yellow triangles)) as shown in Figure 3 (c) and (d). These results indicate that there is only one *sf* minimizing *Er* for each material and the optimal *sf* is material-dependent, which suggests that an improvement can be achieved by making $sf = sf(\boldsymbol{x})$.

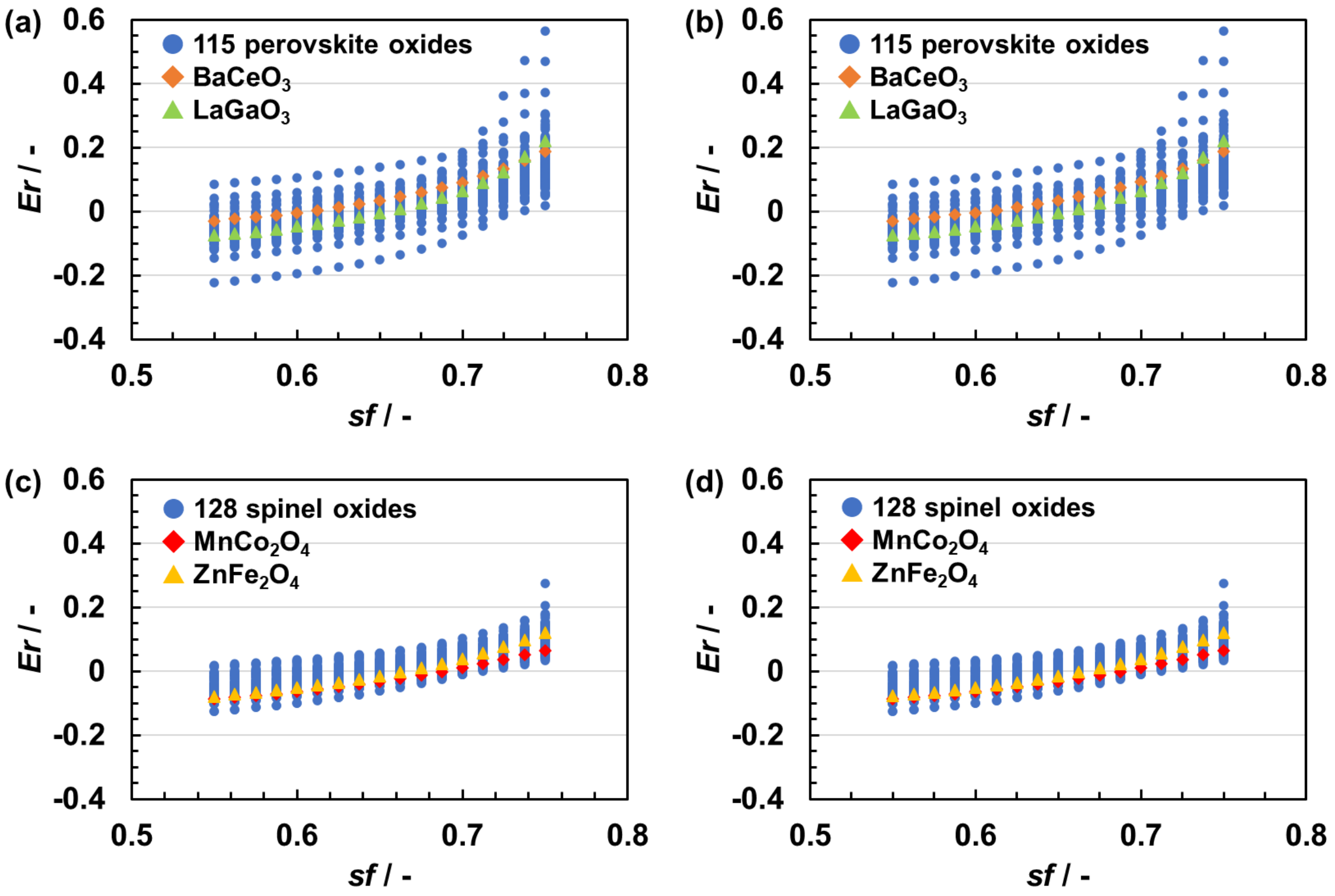

Figure 3. Error in structural parameters (*Er*) of (a, b) perovskite-type oxides and (c, d) spinel-type oxides following SoftBV optimization with different screening factors (*sf*). Figures (a) and (c) are the results of optimizing crystal structures with lattice vectors isotropically contracted by 10%. Figures (b) and (d) are the results of optimizing crystal structures with lattice vectors isotropically expanded by 10%.

The linear and single-hidden layer NN regressions of *sf* for perovskite- and spinel-type oxides were carried out 100 times using different combinations of training and testing data. Figure 4 shows the distributions of root mean square error (RMSE) values of estimated *sf* from these regressions.

Table 1 summarizes the maximum, minimum, and median RMSE and $R^2$ values over the 100 runs. The RMSE for the training data decreases with an increase in the number of nodes for the NN regression, as expected, while the median RMSE for the testing data was the lowest for the NN regressions with only 1 – 3 nodes. The results did not change when the number of the hidden layers changed to 2 – 12. The NN regressions with 1 – 3 nodes show smaller median RMSE for both training and testing data than the linear regression. Therefore, the non-linearity or coupling effects present in an NN might improve the accuracy, which is analyzed in 3.2, but the small number of data makes it difficult.

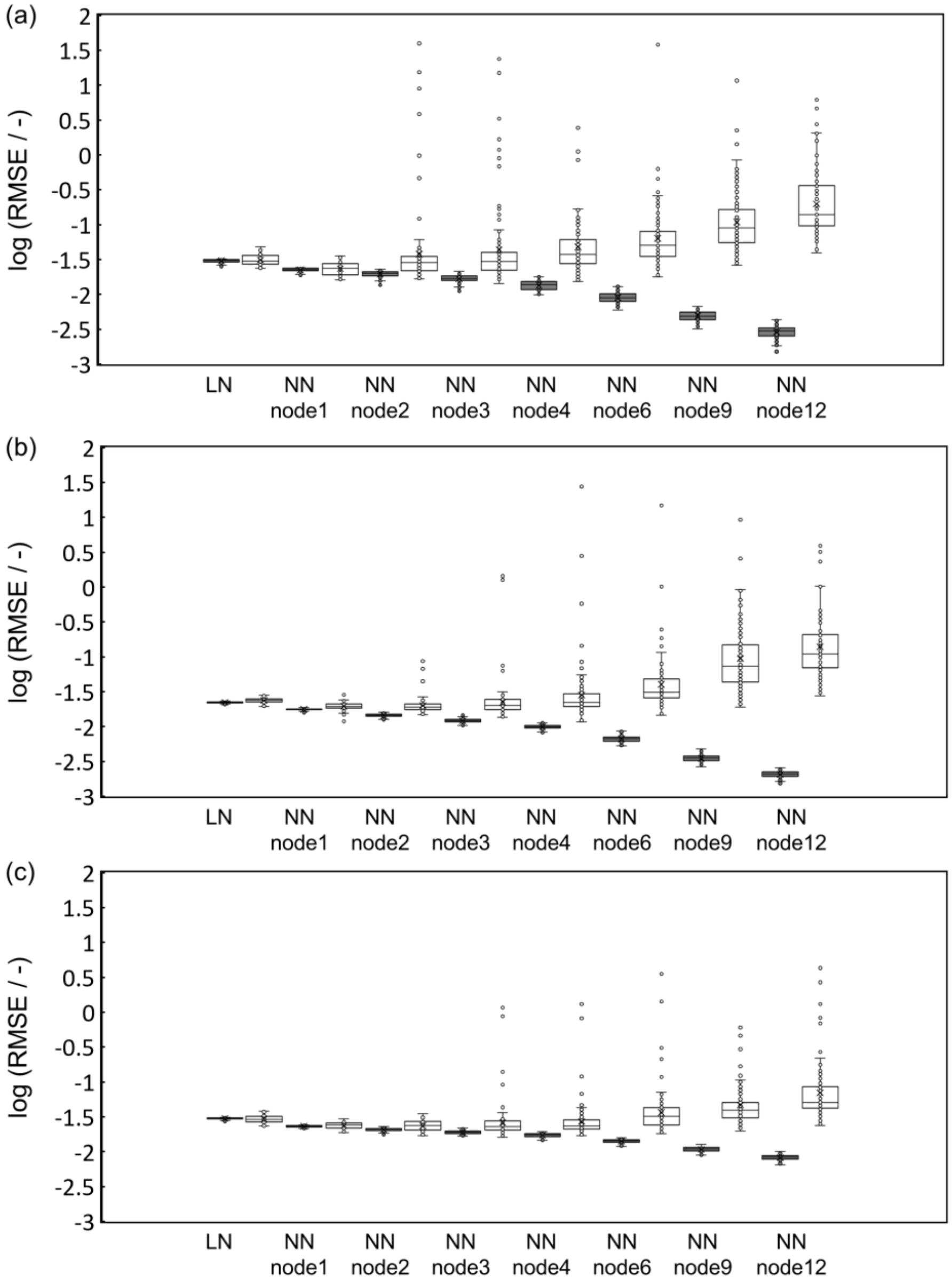

Figure 4. Root mean square error (RMSE) of the screening factor for training (black) and testing (white) data obtained by linear (LN) and neural network (NN) regressions with 1 to 12 nonlinear nodes, for 100 runs with different combinations of the training and testing data for (a) perovskite-type oxides, (b) spinel-type oxides, and (c) the combined dataset.

Table 1. Maximum, minimum, and median root mean square errors (RMSE) and $R^2$ values in the screening factor of linear and neural network (NN) regressions. "$N$" is the number of nodes (neurons).

| Perovskite | | | | | | |
|---|---|---|---|---|---|---|
| Methods | RMSE/$R^2$ of training | | | RMSE/$R^2$ of testing | | |
| | Maximum | Minimum | Median | Maximum | Minimum | Median |
| Linear | 0.032/0.83 | 0.025/0.73 | 0.031/0.75 | 0.049/0.89 | 0.024/0.56 | 0.03/0.76 |
| NN, $N$ = 1 | 0.024/0.90 | 0.019/0.84 | 0.023/0.86 | 0.036/0.93 | 0.016/0.71 | 0.024/0.85 |
| NN, $N$ = 2 | 0.023/0.95 | 0.014/0.86 | 0.02/0.89 | 40/0.93 | 0.017/0.00 | 0.029/0.80 |
| NN, $N$ = 3 | 0.021/0.97 | 0.011/0.88 | 0.017/0.92 | 24/0.95 | 0.014/0.00 | 0.029/0.79 |
| NN, $N$ = 4 | 0.018/0.97 | 0.010/0.91 | 0.014/0.95 | 2.5/0.94 | 0.015/0.01 | 0.036/0.71 |
| NN, $N$ = 6 | 0.013/0.99 | 0.006/0.95 | 0.009/0.98 | 38/0.91 | 0.018/0.00 | 0.051/0.60 |
| NN, $N$ = 9 | 0.007/1.00 | 0.003/0.99 | 0.005/0.99 | 12/0.85 | 0.026/0.00 | 0.088/0.34 |
| NN, $N$ = 12 | 0.004/1.00 | 0.002/0.99 | 0.003/1.00 | 6.2/0.76 | 0.040/0.00 | 0.14/0.17 |
| Spinel | | | | | | |
| Methods | RMSE/$R^2$ of training | | | RMSE/$R^2$ of testing | | |
| | Maximum | Minimum | Median | Maximum | Minimum | Median |
| Linear | 0.023/0.88 | 0.021/0.86 | 0.022/0.87 | 0.028/0.90 | 0.02/0.79 | 0.024/0.85 |
| NN, $N$ = 1 | 0.019/0.93 | 0.016/0.90 | 0.018/0.92 | 0.029/0.96 | 0.012/0.80 | 0.02/0.90 |
| NN, $N$ = 2 | 0.016/0.96 | 0.013/0.93 | 0.015/0.94 | 0.087/0.94 | 0.015/0.33 | 0.019/0.91 |
| NN, $N$ = 3 | 0.015/0.97 | 0.010/0.94 | 0.012/0.96 | 1.4/0.95 | 0.014/0.00 | 0.02/0.90 |
| NN, $N$ = 4 | 0.013/0.98 | 0.008/0.96 | 0.010/0.97 | 28/0.96 | 0.012/0.00 | 0.022/0.88 |
| NN, $N$ = 6 | 0.009/0.99 | 0.005/0.98 | 0.007/0.99 | 15/0.95 | 0.015/0.00 | 0.031/0.80 |
| NN, $N$ = 9 | 0.005/1.00 | 0.003/0.99 | 0.004/1.00 | 9.2/0.91 | 0.019/0.00 | 0.073/0.41 |
| NN, $N$ = 12 | 0.003/1.00 | 0.002/1.00 | 0.002/1.00 | 3.9/0.83 | 0.028/0.00 | 0.11/0.23 |
| Perovskite + Spinel | | | | | | |

| Methods | RMSE/$R^2$ of training | | | RMSE/$R^2$ of testing | | |
|---|---|---|---|---|---|---|
| | Maximum | Minimum | Median | Maximum | Minimum | Median |
| Linear | 0.031/0.80 | 0.027/0.74 | 0.030/0.76 | 0.038/0.86 | 0.024/0.63 | 0.029/0.77 |
| NN, $N$ = 1 | 0.024/0.88 | 0.021/0.84 | 0.023/0.86 | 0.030/0.91 | 0.019/0.77 | 0.024/0.85 |
| NN, $N$ = 2 | 0.023/0.92 | 0.018/0.86 | 0.021/0.88 | 0.035/0.92 | 0.017/0.72 | 0.024/0.86 |
| NN, $N$ = 3 | 0.022/0.92 | 0.017/0.87 | 0.019/0.90 | 1.2/0.93 | 0.016/0.01 | 0.023/0.86 |
| NN, $N$ = 4 | 0.019/0.94 | 0.014/0.90 | 0.017/0.92 | 1.3/0.92 | 0.017/0.00 | 0.024/0.86 |
| NN, $N$ = 6 | 0.016/0.96 | 0.012/0.93 | 0.014/0.95 | 3.5/0.93 | 0.018/0.00 | 0.031/0.77 |
| NN, $N$ = 9 | 0.013/0.98 | 0.009/0.96 | 0.011/0.97 | 0.62/0.90 | 0.020/0.00 | 0.039/0.70 |
| NN, $N$ = 12 | 0.01/0.99 | 0.007/0.97 | 0.008/0.98 | 4.3/0.86 | 0.024/0.00 | 0.051/0.59 |

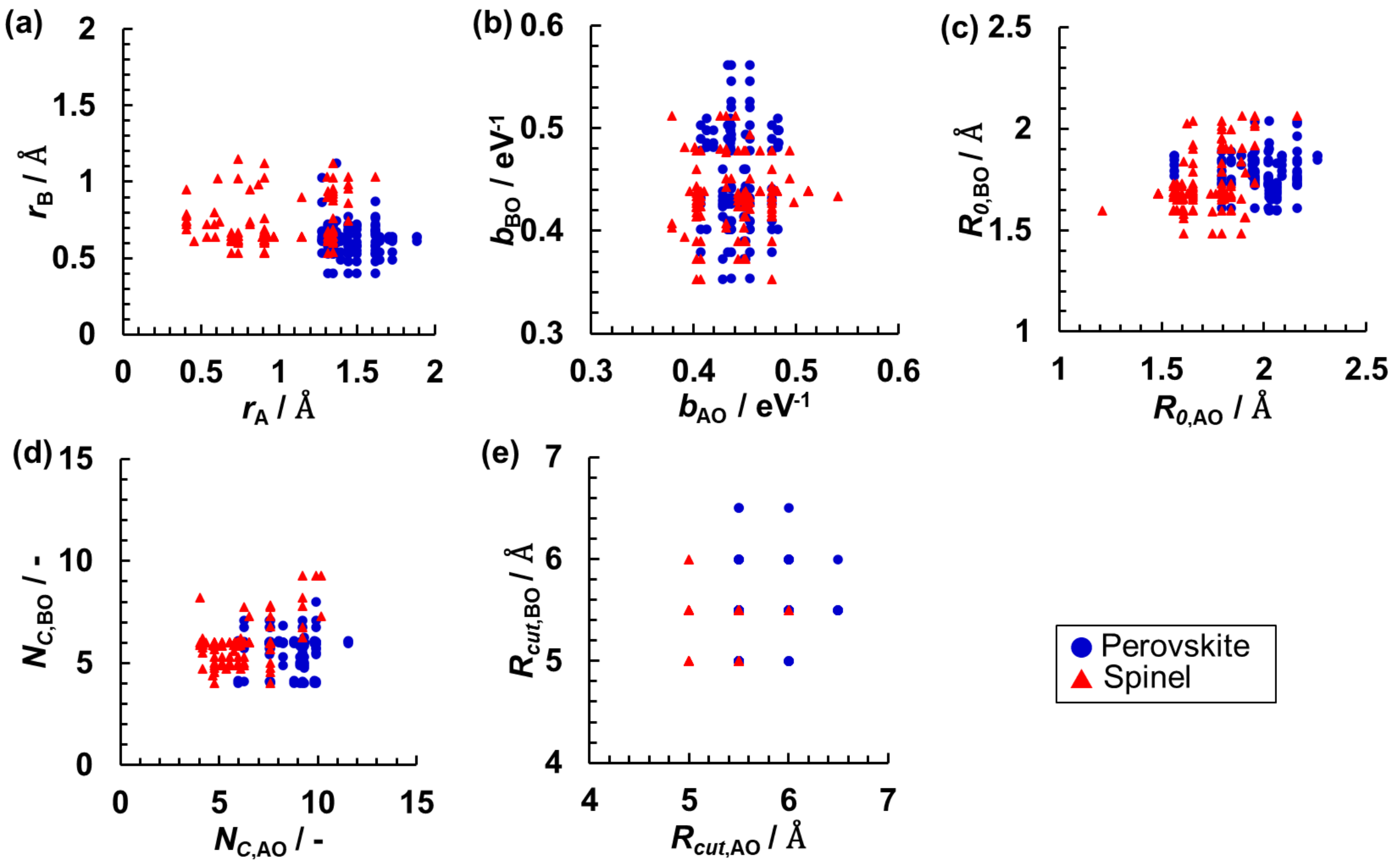

Figure 5. Distributions of pairs of parameters (a: $r_A$ and $r_B$, b: $b_{AO}$ and $b_{BO}$, c: $R_{0, AO}$ and $R_{0, BO}$, d: $N_{C, AO}$ and $N_{C, BO}$, and e: $R_{cut, AO}$ and $R_{cut, BO}$) in perovskite- (blue circles) and spinel-type (red triangles) oxides data.

The RMSE for the testing set could be decreased, especially for the NN regression with a larger number of nodes, by increasing the number of data using both the perovskite- and spinel-type oxides data in a combined dataset. These results show that a key issue is overfitting due to the small number of data points. Figure 5 shows the distributions of the data for selected pairs of parameters (among $b_{ij}, R_{0,ij}, R_{cutoff,ij}, r_i, N_C$). Even from two-dimensional projections that allow only a limited insight into a multivariate distribution, one can appreciate rather uneven and sparse sampling with data based on individual crystal structure types. This result indicates that the accuracy of SoftBV can be improved by estimating *sf* as a function of SoftBV parameters encoding composition if the space of descriptors can be adequately sampled using data for oxides of various compositions.

The crystal structures were optimized using each of the average $sf_{opt}$ computed from each of the five linear, NN, and GPR-NN (shown in 3.2) regression models that had the highest $R^2$ values among the 100 runs (Figure 6). These models used both perovskite- and spinel-type oxide data for training. The use of $sf_{opt}$ improved the accuracy of structure optimization from using $sf_{auto}$. The mean absolute error (MAE) and the standard deviation (STD) of the distributions of *Er* were summarized in Table 2. Although an NN in principle has a higher expressive power and should be able to make a better fit, the MAE and STD for the linear model were equal or even slightly better than the NN model. This ultimately has to do with a small number of data and associated overfitting (see Figure 4). Overall, there is no significant improvement in *sf* fitting quality with NN vs. linear regression, and the NN fit does not lead to an improvement in the estimation of the optimal *sf* and in the quality of structure optimization. While the accuracy has improved on average, the distribution of *Er* with the linear or NN regression is relatively broad with *Er* for some materials exceeding 0.1. The GPR-NN regressions (described in the following section) have the highest accuracy for optimizing the crystal structures with the narrowed distribution of $Er$, with MAE = 0.014 and STD = 0.025.

Figure 7 shows the relationship between GII obtained from the optimized and reference structures. GII is an index for chemical stability, e.g. GII < 0.1 is typically taken to mean that the structure is stable, while GII > 0.2 is considered to be a warning that the structure may be unstable.[21] A better GII value should be obtained when a better structure is used because the error of GII is due to the error of $s_{ij}(R_{ij}) = \exp\left(\frac{R_{0,ij} - R_{ij}}{b_{ij}}\right)$ as Eq. (7), in other words, due to the error in the distance between cations and anions. GII values of optimized structures using $sf_{auto}$ are larger than those of reference structures and do not show the correlation of GII values of SoftBV-optimized structures with those of reference structures. On the other hand, there is a correlation between the GII values of structures optimized using $sf_{opt}$ and the reference structures, especially for perovskite-type oxides. This result reflects the improvement of the accuracy of structure optimization with ML-estimated $sf$.

Table 2. The mean absolute error (MAE) and the standard deviation (STD) for the error of structure optimization of perovskite, spinel, and both oxides using the automatically set screening factors in the SoftBV (“Auto”) and estimated optimal screening factors by the linear, the neural network (“NN”), and the GPR-NN methods trained on the combined data set of the perovskite- and spinel-type oxides.

| | | Auto | Linear | NN (node = 1) | GPR-NN |
|---|---|---|---|---|---|
| Perovskite oxides | MAE | 0.13 | 0.026 | 0.031 | 0.014 |
| | STD | 0.066 | 0.038 | 0.044 | 0.024 |
| Spinel oxides | MAE | 0.10 | 0.023 | 0.022 | 0.013 |
| | STD | 0.032 | 0.024 | 0.026 | 0.026 |
| Both oxides | MAE | 0.12 | 0.025 | 0.026 | 0.014 |
| | STD | 0.053 | 0.032 | 0.036 | 0.025 |

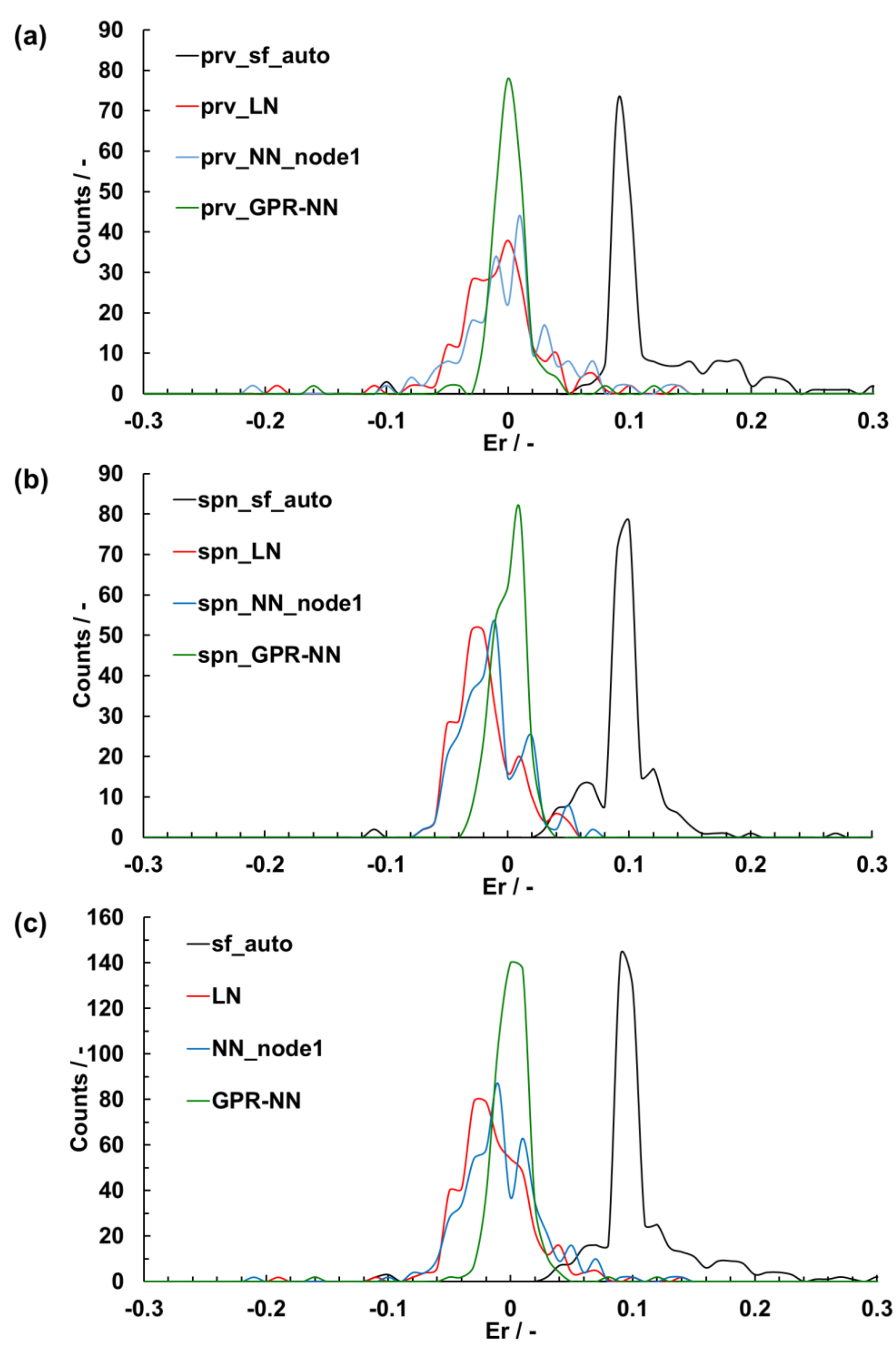

Figure 6. The distribution of structure parameter errors of crystal structures optimized using automatically set screening factors in the SoftBV ("sf_auto") and screening factors estimated by the linear regression ("LN"), neural network with 1

node (“NN_node1”), and the GPR-NN methods trained on the combined data set of the perovskite- and spinel-type oxides, for (a) perovskite (“prv”), (b) spinel (“spn”), and (c) both oxides.

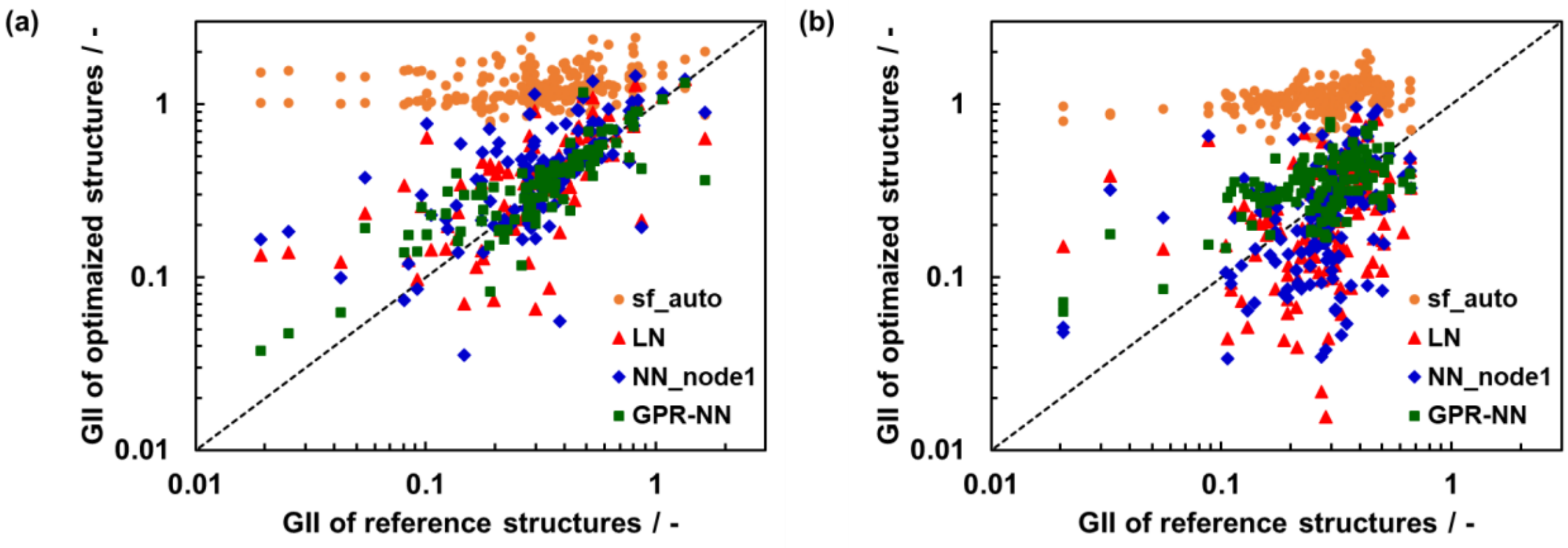

Figure 7. GII of optimized and reference structures of (a) perovskite- and (b) spinel-type oxides. The crystal structures were optimized by using automatically set screening factors in the SoftBV (“sf_auto”) and screening factors estimated by the linear regression (“LN”), the neural network with 1 node (“NN_node1”), and the GPR-NN methods trained on the combined data set of the perovskite- and spinel-type oxides.

### *3.2 Analysis of the importance of nonlinearity and coupling using the GPR-NN method*

The NN results are somewhat unusual in that while there is a slight improvement in the quality of *sf* prediction (judged by the value of $R^2$ over the test set and the range thereof for different train-test splits) over linear regression, there is no improvement in the quality of structure optimization vs. linear regression, and the optimal NN appears to have a size of 1 - 3 neurons only, with the 2- or 3-neuron NN only insignificantly outperforming a 1-neuron NN, with larger NNs showing clear overfitting. NN being a universal approximator, the training set error can be made arbitrarily small, but the global quality of the model, exemplified by the test set error, is ultimately limited by the density of sampling. When sampling is sparse enough, higher-order coupling terms may not be recoverable.[54,58,60] That the sampling is sparse in this case, and that this is a limiting factor in utilizing the superior expressive power of an NN, is clear from the above comparison of fitting only the perovskite or the spinel data separately or the combined dataset.

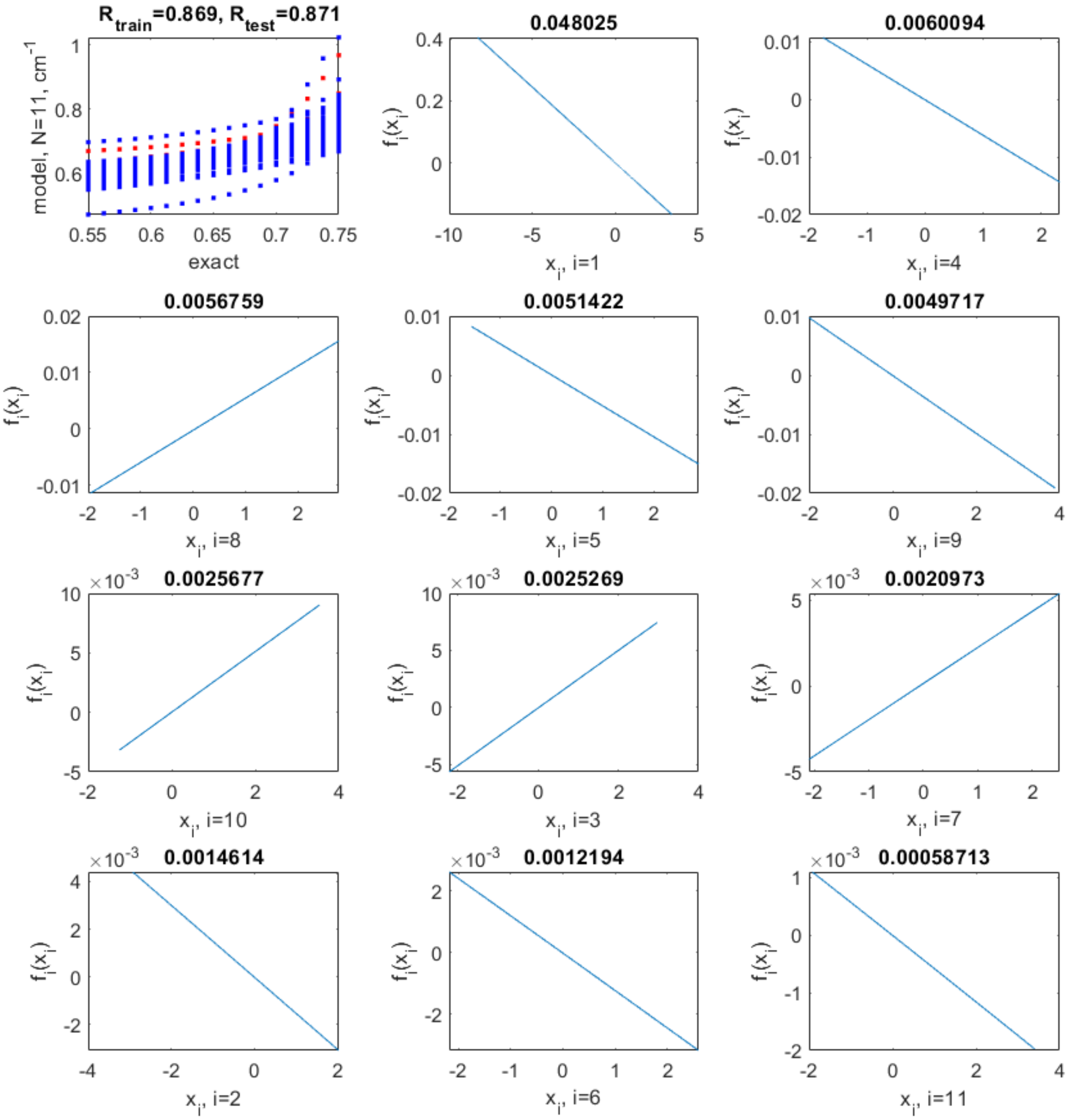

Figure 8. Top left: correlation between target ("exact") values of the screening factor and those predicted by an additive model with a kernel length set to a large value $l$ = 200, for training (blue) and test (red) data (some blue and red points

visually overlap). The correlation coefficients between the exact and predicted values for training and testing data are also shown. The following panels show the shapes of $f_i(x_i)$ in the order of decaying magnitude, with the magnitude (defined as $var(f_i)^{1/2}$) shown on top of each plot.

A NN performs non-linear operations on linear combinations of inputs $\{x_n\}$ introducing both nonlinearity and coupling. This is true even for a single-hidden neuron NN. We can separate these two effects with the help of the GPR-NN method. We first perform simulations where $\boldsymbol{y}$ = $\boldsymbol{x}$, i.e. an additive model in $\boldsymbol{x}$, $sf(\boldsymbol{x}) = \sum_{n=1}^{N} f_n(x_n)$. We perform a two-dimensional hyperparameters scan of the length parameter $l$ and the GPR noise parameter $\sigma$. At each $(l, \sigma)$, we perform 100 fits differing by different random splits of training and test data (whereby 20 percent of materials are used for testing and 80 for training). Not that when $l$ becomes large ($l \gg 1$ for data scaled on unit cube), kernel resolution is lost[61] and the component functions $f_n(x_n)$ become near-linear. This is illustrated in Figure 8 for the case of $l$ = 200, $\log(\sigma) = -3$, where we show the shapes of $f_n$ in such a limiting case as well as the correlation plots between the exact (target) values of *sf* and those predicted by the model for a representative run. In this case the average/min/max/standard deviation (over 100 runs) of the training set $R^2$ are 0.80/0.78/0.84/0.02, and of the test set $R^2$, 0.77/0.59/0.85/0.06, respectively, - similar to traditional linear regression. The average/min/max/standard deviation of the RMSE is 0.031/0.028/0.032/0.001 for the training and 0.032/0.028/0.039/0.002 for the test set, respectively.

The optimal hyperparameters were chosen as those minimizing simultaneously the average test set $R^2$ and its variance (over multiple runs); they are $l$ = 7 and $\log(\sigma) = -3$. With these hyperparameters, the average/min/max/standard deviation (over 100 runs) of the training set $R^2$ are 0.89/0.88/0.92/0.01, respectively, and of the test set $R^2$, 0.85/0.71/0.93/0.05, respectively. The average/min/max/standard deviation of the RMSE is 0.022/0.019/0.023/0.001 for the training and 0.024/0.017/0.038/0.006 for the test set, respectively. This is a noticeable improvement over linear regression and the NN. This model has no coupling. The correlation plots between the exact (target) values of *sf* and those predicted by the model as well as the shapes of $f_n$ in this case are shown in Figure 9 for a representative run. They are highly nonlinear. Nonlinearity improves the quality of the model and also influences the relative importance of variables: in both the linear and the nonlinear model, the most important (by the magnitude of $f_n(x_n)$) variables are $x_1$ (*Er*), $x_4$ ($R_{0,AO}$), and $x_8$ ($r_A$). The least important is $x_{10}$ ($N_{C,AO}$) in the non-linear model with the optimal $l$ = 7 while it is $x_{11}$ ($N_{C,BO}$) in the (practically) linear model achieved with $l$ = 200. The order of importance of variables with small magnitudes of $f_n(x_n)$ may differ; it is normal that the relative importance of features is different for different methods.[62,63]

We now fix $l$ and σ at their optimized values and test if adding coupling terms further improves the model. The results are summarized in Figure 10. We do not observe any further improvement due to the inclusion of coupling among the features. The coupling terms are either unimportant or unrecoverable due to the low density of sampling.

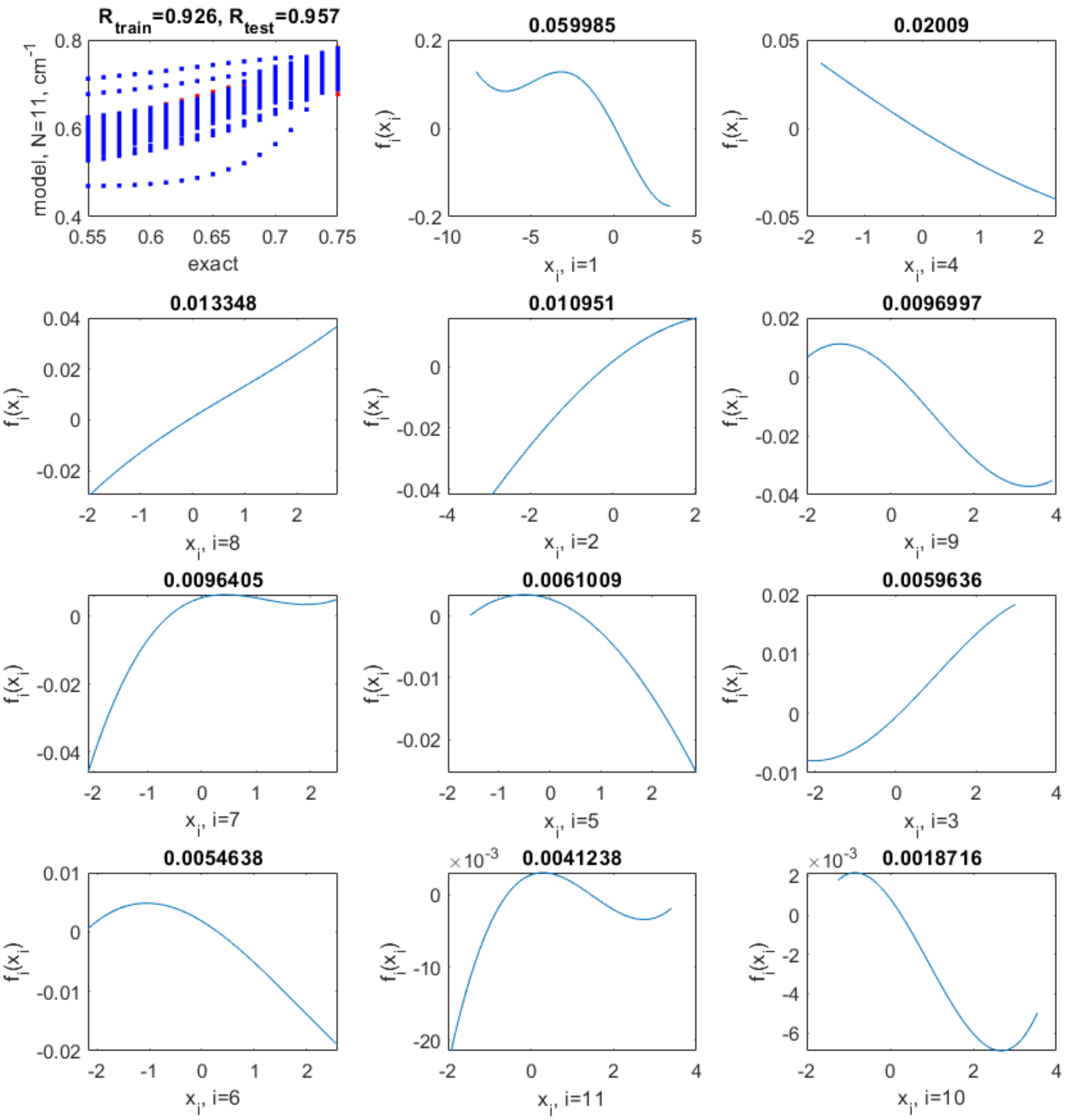

Figure 9. Top left: correlation between target ("exact") values of the screening factor and those predicted by an additive model with an optimized kernel length of $l$ = 7, for training (blue) and test (red) data (some blue and red points visually overlap). The correlation coefficients between the exact and predicted values for training and test data are also shown. The following panels show the shapes of $f_i(x_i)$ in the order of decaying magnitude, with the magnitude (defined as $var(f_i)^{1/2}$) shown on top of each plot.

Finally, in Figure 6, we show the distribution of structural parameter errors achieved with the GPR-NN method (using optimal hyperparameters). The method is clearly superior over the linear regression and the NN in terms of the average error as well as the width of the error distribution, which are listed in Table 2. The optimal shapes of the nonlinear functions used with each variable, and the absence of nonlinear parameter optimization in GPR-NN allow capitalizing on the superior expressive power of a nonlinear method while retaining the robustness of linear regression.

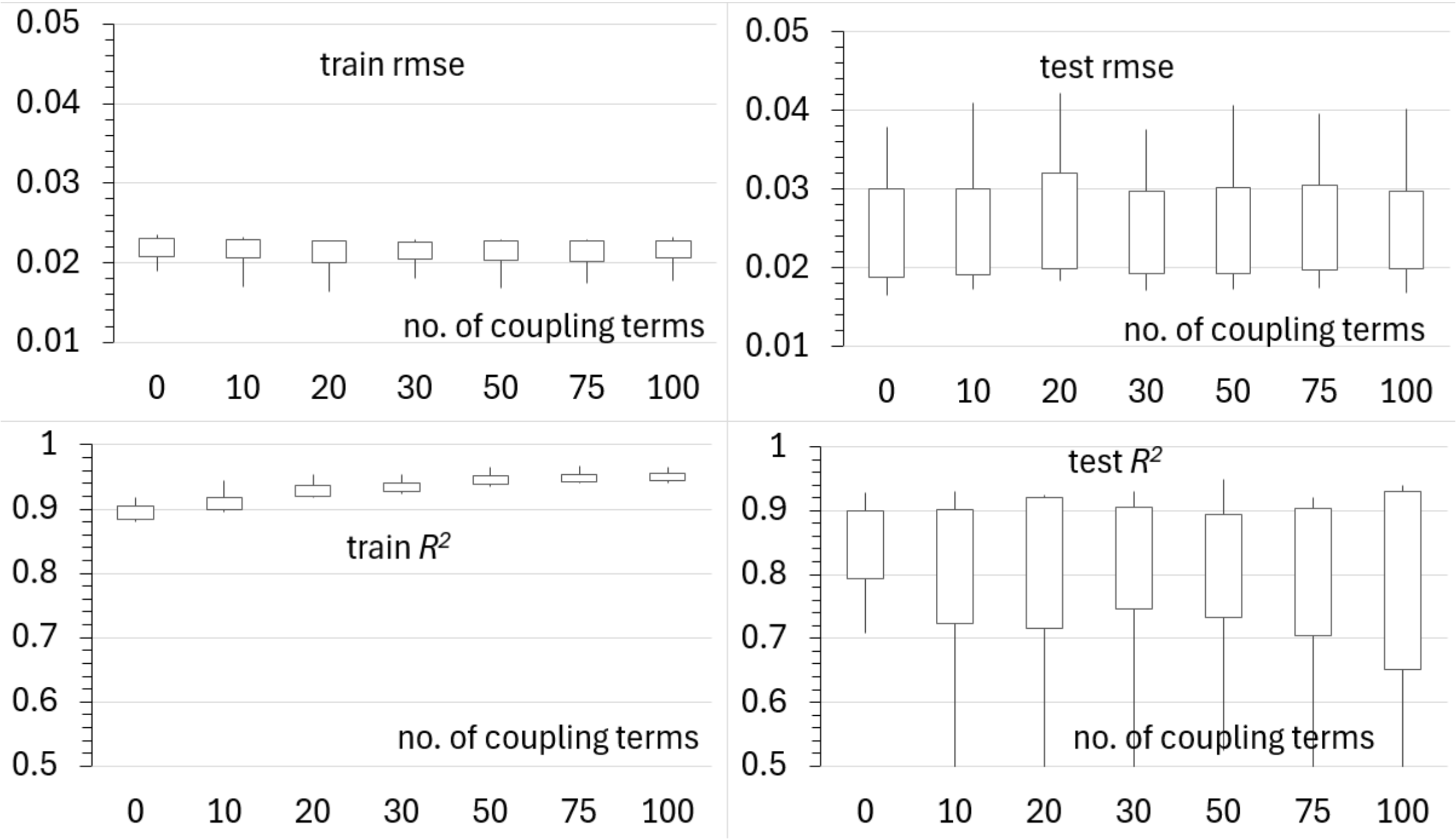

Figure 10. Statistics of training and test set errors in *sf* and $R^2$ values as a function of the number of coupling terms. The box shows a one-sigma interval about the mean, and the whiskers show minimum and maximum values, over 100 runs differing by random selection of training and test data.

# 4 Conclusions

In this study, we explored the possibility and extent of improvement of the accuracy of the SoftBV approximation by fitting the screening factor as a function of descriptors of chemical composition. We showed that it is the screening factor that can be parameterized in this way without the danger of tempering with the basis of SoftBV ideology. The features that we used are various parameters that are already available in a SoftBV calculation; that is, the screening factor as a function of those features can in principle be implemented without hardship. We first used linear and neural network models and showed, on the examples of perovskite- and spinel-type oxides which have been proposed as promising solid-state ionic conductors, that this can noticeably improve the ability of the SoftBV approximation to model structures, in particular new, putative crystal structures whose structural parameters are yet unknown.

We showed that the sampling density of the space of descriptors is an important limiting factor in the possible improvement in *sf*, which may even prevent one from using the superior expressive power of nonlinear models. In this work, this was palliated on one hand by combining data from different crystal structures having structural similarity (perovskite and spinel oxides in this case) and on the other hand by producing synthetic sample points from strained structures. Only a slight improvement in the screening factor regression was obtained with an NN over linear regression while no improvement over linear regression was observed in the quality of structure optimization with *sf* predicted by the NN model.

We then applied to this problem the recently developed GPR-NN method that allows obtaining a superior expressive power of a nonlinear approximation while avoiding nonlinear parameter optimization during regression. The method is a hybrid between an NN and kernel regression; it builds optimal shapes of nonlinear basis functions (neuron activation functions) and permits including coupling among features in a controlled way. We analyzed the relative importance of nonlinearity and coupling and found that while nonlinearity helps obtain a more accurate model, coupling terms were not important or were unrecoverable from the data. The *sf* predicted by GPR-NN showed the best quality of structure optimization with SoftBV and a significant improvement over linear and NN regressions.

## 5 Declaration of Competing Interests

We declare that we have no conflict of interest.

## 6 Author contributions

Keisuke Kameda: Data curation, Software, Visualization, Investigation, Formal Analysis, Writing- Original draft preparation, Writing- Reviewing and Editing

Takaaki Ariga: Data curation, Software, Visualization, Investigation, Formal Analysis

Kazuma Ito: Data curation, Software, Formal Analysis

Manabu Ihara: Supervision, Project Administration, Funding Acquisition, Resources, Writing- Reviewing and Editing

Sergei Manzhos: Conceptualization, Methodology, Supervision, Resources, Project Administration Writing- Original draft preparation, Writing- Reviewing and Editing

## 7 Supplementary Material

Supplementary Material contains the list of structures used in this work together with their respective database identifiers, as well as the dataset used for machine learning.

## 8 Acknowledgments

This work was supported by JST-Mirai Program, Japan, Grant Number JPMJMI22H1. We thank Prof. Stefan Adams of National University of Singapore for the discussions and consultations. We thank Digital Research Alliance of Canada on whose computers some of the calculations were performed.

## 9 Data availability

A list of all crystal structures with their database identifiers as well as the dataset used in machine learning are available in Supplementary Materials. The GPR-NN code is available in Ref. [45].

# Supplementary material

# Machine learning the screening factor in the soft bond valence approach for rapid prescreening of ceramics

Keisuke Kamda, Takaaki Ariga, Kazuma Ito, Manabu Ihara[1] Sergei Manzhos[2],

School of Materials and Chemical Technology, Tokyo Institute of Technology, Ookayama 2-12-1, Meguro-ku, Tokyo 152-8552 Japan.

## List of perovskite structures used

Structures taken from ICDD database [1] are marked with *. All other structures were taken from Materials Project database [2].

| | | | |
|---|---|---|---|
| BaCeO3 | mp-5663 | LaGaO3 | mp-1097026 |
| BaCoO3 | mp-1076782 | LaMnO3 | mp-19025 |
| BaCrO3* | #04-022-1253 | LaNiO3 | mp-1075921 |
| BaFeO3 | mp-19035 | LaScO3 | mp-1096800 |
| BaHfO3 | mp-998552 | LaTiO3 | mp-8020 |
| BaMnO3 | mp-1016852 | LaVO3 | mp-19053 |

[1] E-mail: mihara@chemeng.titech.ac.jp
[2] E-mail: manzhos.s.aa@m.titech.ac.jp

| | | | |
|---|---|---|---|
| BaMoO3 | mp-19322 | NaCrO3 | mp-1076642 |
| BaNbO3 | mp-3020 | NaMoO3 | mp-1040471 |
| BaNiO3 | mp-1120765 | NaNbO3 | mp-3136 |
| BaPbO3 | mp-21280 | NaTaO3 | mp-4170 |
| BaSiO3 | mp-1016821 | NaVO3 | mp-1099591 |
| BaSnO3 | mp-3163 | NaWO3 | mp-19328 |
| BaTaO3 | mp-754678 | NdAlO3 | mp-14254 |
| BaTbO3 | mp-2929 | NdBiO3 | mp-974740 |
| BaTiO3 | mp-2998 | NdCoO3 | mp-20031 |
| BaVO3 | mp-1017465 | NdCrO3 | mp-19062 |
| BaWO3 | mp-1183395 | NdGaO3 | mp-9834 |
| BaZrO3 | mp-3834 | NdInO3 | mp-1186316 |
| CaCoO3 | mp-1099934 | NdTiO3* | #04-002-3783 |
| CaFeO3 | mp-1001571 | NdVO3 | mp-19253 |
| CaHfO3 | mp-1016873 | NdYbO3 | mp-1187576 |
| CaMnO3 | mp-1017467 | PbCrO3 | mp-22364 |
| CaSiO3 | mp-5893 | PbFeO3 | mp-973579 |
| CaSnO3 | mp-7986 | PbHfO3 | mp-22535 |
| CaTiO3 | mp-5827 | PbMnO3 | mp-37214 |
| CaVO3 | mp-1016853 | PbMoO3 | mp-1186106 |
| CaZrO3 | mp-542112 | PbNiO3 | mp-974108 |
| CdHfO3 | mp-1017446 | PbSiO3 | mp-978489 |
| CdMnO3 | mp-1016854 | PbSnO3 | mp-978952 |
| CdSiO3 | mp-1016879 | PbTiO3 | mp-19845 |
| CdSnO3 | mp-1016881 | PbVO3 | mp-1070440 |
| CdTiO3 | mp-22345 | PbZrO3 | mp-1068577 |
| CdVO3 | mp-1016904 | RbCrO3 | mp-1076360 |
| CdZrO3 | mp-1016845 | RbMoO3 | mp-975292 |
| CeAlO3 | mp-5323 | RbNbO3 | mp-1075911 |
| CeCrO3 | mp-20530 | RbTaO3 | mp-1076534 |
| CeCuO3 | mp-977389 | RbVO3 | mp-1076638 |
| CeFeO3 | mp-864636 | RbWO3 | mp-975138 |
| CeGaO3 | mp-33365 | SrCoO3 | mp-505766 |
| CeMnO3 | mp-1183706 | SrCrO3 | mp-20029 |
| CeNiO3 | mp-866095 | SrFeO3 | mp-510624 |
| CeTiO3 | mp-754524 | SrHfO3 | mp-4551 |
| CeVO3 | mp-22593 | SrMnO3 | mp-1017466 |
| CsMoO3 | mp-1183917 | SrMoO3 | mp-18747 |
| CsNbO3 | mp-1096944 | SrNbO3 | mp-7006 |
| CsTaO3 | mp-1185552 | SrNiO3 | mp-762506 |
| KCrO3 | mp-1076732 | SrPbO3* | #04-008-0331 |
| KMoO3 | mp-1040469 | SrSiO3 | mp-1017439 |
| KNbO3 | mp-935811 | SrSnO3 | mp-546973 |
| KTaO3 | mp-3614 | SrTaO3 | mp-1186755 |
| KVO3 | mp-1076633 | SrTiO3 | mp-5229 |
| KWO3 | mp-1040472 | SrVO3 | mp-18717 |
| LaAgO3 | mp-1076000 | SrWO3 | mp-1186764 |
| LaAlO3 | mp-5304 | SrZrO3 | mp-613402 |
| LaCoO3 | mp-573180 | TlNbO3 | mp-977408 |
| LaCrO3 | mp-18841 | TlTaO3 | mp-861873 |
| LaCuO3 | mp-1076070 | TlWO3 | mp-1187621 |
| LaFeO3 | mp-552676 | | |

## List of spinel structures used

| | | | |
|---|---|---|---|
| BaLa2O4 | mp-755558 | MgCu2O4 | mvc-4609 |
| BeCo2O4 | mp-770957 | MgFe2O4 | mp-608016 |
| CaAg2O4 | mvc-4692 | MgGa2O4 | mp-4590 |
| CaBi2O4 | mvc-4662 | MgIn2O4 | mp-7831 |
| CaCo2O4 | mvc-11995 | MgMn2O4 | mvc-15009 |
| CaCr2O4 | mp-1304962 | MgMo2O4 | mvc-4795 |
| CaCu2O4 | mvc-4685 | MgNi2O4 | mp-1319349 |
| CaFe2O4 | mvc-13150 | MgRh2O4 | mp-3319 |
| CaGd2O4 | mp-752679 | MgSb2O4 | mvc-4678 |
| CaIn2O4 | mp-22766 | MgTi2O4 | mp-27872 |
| CaMo2O4 | mp-1539672 | MgV2O4 | mp-18900 |
| CaNi2O4 | mp-1273583 | MnAl2O4 | mp-755882 |
| CaSb2O4 | mvc-4658 | MnCo2O4 | mp-1222025 |
| CaSm2O4 | mp-754240 | MnCr2O4 | mp-28226 |
| CaTb2O4 | mp-755044 | MnFe2O4 | mp-18750 |
| CaTi2O4 | mvc-6014 | MnIn2O4 | mp-35162 |
| CaTm2O4 | mp-1178472 | MnRh2O4 | mp-554354 |
| CaV2O4 | mvc-11563 | MnTi2O4 | mp-561097 |
| CaY2O4 | mp-753815 | MnV2O4 | mp-35475 |
| CdAl2O4 | mp-36866 | MoAg2O4 | mp-19318 |
| CdCo2O4 | mp-756301 | MoNa2O4 | mp-18852 |
| CdCr2O4 | mp-19262 | NiAl2O4 | mp-688785 |
| CdFe2O4 | mp-21333 | NiCo2O4 | mp-1096547 |
| CdGa2O4 | mp-3443 | NiCr2O4 | mp-19303 |
| CdGd2O4 | mp-754093 | NiFe2O4 | mp-22684 |
| CdIn2O4 | mp-19803 | NiGa2O4 | mp-756649 |
| CdRh2O4 | mp-14100 | NiMn2O4 | mp-29399 |
| CdV2O4 | mp-18847 | NiRh2O4 | mp-19307 |
| CoAl2O4 | mp-36447 | PdNd2O4 | mp-1210248 |
| CoCr2O4 | mp-20758 | PdZn2O4 | mp-22257 |
| CoFe2O4 | mp-753222 | SiCd2O4 | mp-560842 |
| CoGa2O4 | mp-765466 | SiCo2O4 | mp-19071 |
| CoMg2O4 | mp-753991 | SiFe2O4 | mp-18816 |
| CoNi2O4 | mp-754168 | SiMg2O4 | mp-5639 |
| CoRh2O4 | mp-546936 | SiNi2O4 | mp-18766 |
| CoV2O4 | mp-758452 | SiV2O4 | mp-754234 |
| CuAl2O4 | mp-27719 | SiZn2O4 | mp-558096 |
| CuCo2O4 | mp-34146 | SnCd2O4 | mp-1104726 |
| CuCr2O4 | mp-504573 | SnMg2O4 | mp-973261 |
| CuFe2O4 | mp-770107 | SnZn2O4 | mp-1103830 |
| CuGa2O4 | mp-753397 | SrLa2O4 | mp-754211 |
| CuMn2O4 | mp-505421 | SrLu2O4 | mp-756646 |
| CuNi2O4 | mp-756271 | SrNd2O4 | mp-753418 |
| CuRh2O4 | mp-4409 | SrSc2O4 | mp-754114 |
| EuLa2O4 | mp-1178267 | SrSm2O4 | mp-754942 |
| EuY2O4 | mp-754557 | VCr2O4 | mp-754077 |
| FeAl2O4 | mp-30084 | VMg2O4 | mp-30545 |
| FeCr2O4 | mp-20168 | WNa2O4 | mp-18803 |
| FeMg2O4 | mp-768465 | ZnAg2O4 | mvc-4660 |
| FeNi2O4 | mp-640147 | ZnAl2O4 | mp-2908 |
| FeV2O4 | mp-20167 | ZnBi2O4 | mvc-4703 |
| GeMg2O4 | mp-3904 | ZnCo2O4 | mp-753489 |
| HgAl2O4 | mp-756317 | ZnCr2O4 | mp-19410 |
| HgCo2O4 | mp-754069 | ZnCu2O4 | mvc-4675 |
| HgCr2O4 | mp-21074 | ZnFe2O4 | mp-19313 |
| HgFe2O4 | mp-754491 | ZnGa2O4 | mp-5794 |
| HgGa2O4 | mp-755239 | ZnIn2O4 | mp-756297 |
| HgIn2O4 | mp-753983 | ZnMn2O4 | mvc-11612 |
| HgY2O4 | mp-755634 | ZnMo2O4 | mvc-4829 |
| MgAg2O4 | mvc-4630 | ZnNi2O4 | mp-768586 |
| MgAl2O4 | mp-3536 | ZnRh2O4 | mp-5146 |
| MgBi2O4 | mvc-4682 | ZnSb2O4 | mvc-4661 |
| MgCo2O4 | mp-756442 | ZnTi2O4 | mvc-5983 |
| MgCr2O4 | mp-19202 | ZnV2O4 | mp-18879 |